\documentclass[aps,pra,showpacs,twocolumn,floatfix]{revtex4}
\usepackage{float,subfig,amsmath,amsfonts,amsthm,amssymb,bm}
\usepackage{url,enumerate}
\usepackage{epsfig,graphicx}
\newtheorem{theorem}{Theorem}[section]
\usepackage{color}

\begin{document}
\title{Nonlinear Scattering of a Bose-Einstein Condensate on a Rectangular Barrier}
\author{Lincoln D. Carr$^{1,2}$}
\author{Rachel R. Miller$^{1,3}$}
\author{Daniel R. Bolton$^{1,4}$}
\author{{Scott A. Strong$^{1}$}}
\address{$^1$Department of Physics, Colorado School of Mines, Golden,
  Colorado 80401, USA}
\address{$^2$Physikalisches Institut, Universit\"at Heidelberg,
Im Neuenheimer Feld 226, D-69120 Heidelberg, Germany}
\address{$^3$Department of Mathematical and Computer Sciences, Colorado
  School of Mines, Golden, Colorado 80401, USA}
\address{$^4$Department of Physics, University of Washington, Seattle,
  Washington 98195-1560, USA}
\date{\today}

\begin{abstract} We consider the nonlinear scattering and
transmission of an atom laser, or Bose-Einstein condensate
(BEC) on a finite rectangular potential barrier.  The
nonlinearity inherent in this problem leads to several new
physical features beyond the well-known picture from
single-particle quantum mechanics.  We find numerical
evidence for a denumerably infinite string of bifurcations
in the transmission resonances as a function of nonlinearity
and chemical potential, when the potential barrier is wide
compared to the wavelength of oscillations in the
condensate.  Near the bifurcations, we observe extended
regions of near-perfect resonance, in which the barrier is
effectively invisible to the BEC.  Unlike in the linear
case, it is mainly the barrier width, not the height, that
controls the transmission behavior.  We show that the
potential barrier can be used to create and localize a dark
soliton or dark soliton train from a phonon-like standing
wave.  \end{abstract}


\maketitle

\section{Introduction}\label{sec:introduction}

Bose-Einstein condensates (BECs) in the dilute gas limit and
in the presence of many kinds of external potentials are
modeled very well by the nonlinear Schr\"odinger equation
(NLSE)~\cite{dalfovo1999,ketterle1999}.  BECs with potential
barriers have many practical applications, including atom
lasers~\cite{anderson1998,hagley1,bloch2} and
high-resolution holography~\cite{zobay1999,helmerson1}.  Two
recent experiments at Washington State~\cite{engelsP2007}
and Rice~\cite{dries2010} have considered the dynamics of a
finite barrier dragged over an effectively 1D BEC, giving
one the opportunity to investigate the Landau
criterion~\cite{dalfovo1999} under the constraint of a 1D
system.  Instead of the quantized vortices common to
superfluids in two and higher dimensions, dark solitons
appear in 1D.  As the stationary NLSE with a delta function
potential can be solved
exactly~\cite{hakim1,pavloff2002,carr2005a,witthaut2005}, it
is convenient for theory to model localized potential
barriers with delta functions. For instance, multiple delta
functions have been used to investigate intriguing problems
such as nonlinear band structure~\cite{carr2005d} and
Anderson localization~\cite{paulT2007}.  However, as we show
in this article, when the delta function is replaced with a
potential of finite width new features arise, including a
denumerably infinite series of bifurcations in the
transmission resonances, and one or more dark solitons
localized on the barrier.

Since the stationary NLSE can be solved exactly for any
piece-wise constant potential, we take the potential of
finite width to be rectangular in form~\cite{footnote1}.  In
single-particle quantum mechanics, stationary solutions of
the Schr\"odinger equation for a rectangular potential
barrier or well is a classic problem taught in undergraduate
quantum mechanics courses.  The nonlinearity inherent in the
mean-field description of the BEC as captured by the NLSE
substantially changes this well-known problem, as we will
show.  The physical context of our problem is the
steady-state transmission behavior of an atom laser incident
on a barrier.  We assume that the BEC is confined in the
transverse directions by a harmonic oscillator trap, and
that its behavior is
quasi-one-dimensional~\cite{olshanii1998,petrov2000b,carr2000e,dunjko2001};
that is, the longitudinal direction of the BEC is much
larger than the transverse directions and the healing
length, and the chemical potential is much larger than the
transverse excitation energy.  Further, we assume that the
width of the potential barrier is much less than the
longitudinal dimension of the BEC, so that the system is
effectively longitudinally infinite and far-field effects
may be neglected.  This physical situation was
experimentally produced in the Washington State and Rice
experiments~\cite{engelsP2007,dries2010}.

Past studies of the 1D stationary NLSE with various boundary
conditions and potentials are very extensive, including both
repulsive and attractive interactions;
see~\cite{kevrekidisPG2008} and references therein.  In this
article we focus on the more common case of repulsive
interactions, although our techniques work for general
interactions in the NLSE.  Our own work has often treated
cases described by piecewise constant potentials, starting
with a uniform potential with periodic or box boundary
conditions~\cite{carr2000e,carr2000a,carr2000b}.  This work
was later extended to a discontinuous potential step and to
a delta-function potential~\cite{carr2005a}.  Other studies
in this direction included the Kronig-Penney
potential~\cite{carr2005d} and the bichromatic
lattice~\cite{carr2005e}.

The potential barrier or well has been treated previously in
restricted circumstances in three previous studies.  First,
Carr, Mahmud, and Reinhardt~\cite{carr2001f} found
particular classes of bound states in the potential well.
Such bound states are localized or partially localized, and
it was found that experimental parameters could be tuned to
achieve different regimes of tunneling.  Second, Rapedius,
Witthaut and Korsch~\cite{rapedius2006} took the region
outside the well to be strictly linear.  They found a
bistable behavior in the transmitted flux for unbound
states.  The critical nonlinearity at which bound states
appear was discovered.  Third, Ishkhanyan and
Krainov~\cite{ishkhanyan2009} considered the limit of very
small nonlinearity.  They used perturbation theory to expand
the NLSE, and a multi-scale method to find the solutions.
The reflection coefficient $R$ could then be determined by
the usual methods of linear quantum mechanics, decomposing
the wave functions in each region into left- and
right-traveling waves via the superposition principle.  The
latter two studies had in common that they retained the
concept of the superposition principle outside the well.  We
discard this concept in favor of full nonlinearity, solving
the complete nonlinear problem without approximations of any
kind.  Additional related work on scattering of a bright
soliton on a potential barrier or well has treated a variety
of applied and foundational issues in quantum mechanics,
from bound state spectroscopy and resonant
trapping~\cite{ernstT2010} to macroscopic
superpositions~\cite{weissC2009,streltsov2009}
(Schr\"odinger cat states).

In all studies of the stationary 1D NLSE solitons play a key
role.  In the Washington State
experiment~\cite{engelsP2007}, the barrier was produced by
an elliptical laser beam, then dragged through the BEC.  For
intermediate drag speeds a train of dark solitons was
observed in the presence of the barrier.  Such solitons have
been produced by other experimental methods, for example, by
merging two coherent BECs~\cite{denschlag2000}.  An
experiment by Weller \textit{et al}. at Heidelberg studied
the dynamical behavior and interactions of such solitons
\cite{weller2008}.  Solitons are localized, persistent,
robust nonlinear structures which appear often in BEC
experiments \cite{burger1999, carr2002b}.  A key feature of
solitons in quasi-1D in the absence of an external potential
is that they interact elastically and do not
dissipate~\cite{reinhardt1997, stellmerS2008}.  Such
behavior was observed in the Heidelberg experiment, and was
found to agree with numerical simulations of NLSE solution
dynamics~\cite{weller2008}.

Our article is outlined as follows.  In
Sec.~\ref{sec:equation} we present the fundamental equations
and a general solution method.  We emphasize that our work
is completely analytical and symbolic up to one numerical
integration used to characterize transmission; the NLSE
solution itself is analytical and is determined
rigorously~\cite{RRMthesis}.  The appendices contains some
key brief proofs to support this rigor, as well as numerical
studies to support our analysis and a brief overview of
Jacobi elliptic functional relationships used in our
analysis.  In Sec.~\ref{sec:soliton} we apply our solution
method to scattering on the rectangular barrier and obtain
characteristic solutions, highlighting cases for which
solitons are localized on the barrier.  In
Sec.~\ref{sec:transmission} we study transmission of the BEC
for both wide and narrow barriers.  In
Sec.~\ref{sec:resonances} we focus in on the transmission
resonances, where the barrier is invisible, and discover a
series of bifurcations in such resonances. Finally, in
Sec.~\ref{sec:conclusions} we conclude.

\section{Equation and Methods}\label{sec:equation}

\subsection{The 1D Nonlinear Schr\"odinger Equation}

The 1D NLSE is
\begin{multline}
  \label{eq:unitsNLS}
  \left[-\frac{\hbar^2}{2M}\frac{\partial^2}{\partial\tilde{x}^2} + \tilde{g}\vert\tilde{\Psi}(\tilde{x},\tilde{t})\vert^2 +
    \tilde{V}(\tilde{x})\right]\tilde{\Psi}(\tilde{x},\tilde{t})\\ =
  i\hbar\frac{\partial}{\partial\tilde{t}}\tilde{\Psi}(\tilde{x},\tilde{t}),
\end{multline}
where $\tilde{g}=2 a_s \hbar \omega $ is the interaction
strength or nonlinearity renormalized to
1D~\cite{carr2000e}, $a_s$ is the $s$-wave scattering length
for binary contact interactions, $\omega$ is the oscillation
frequency of the transverse trap as described in
Sec.~\ref{sec:introduction}, $M$ is the mass of the atoms or
molecules that are Bose-condensed, and
$\tilde{V}(\tilde{x})$ is an external potential.  We take
the Landau interpretation of the wavefunction or order
parameter: $\tilde{\rho}\equiv |\tilde{\Psi}|^2$ is the
local BEC number density and
$\tilde{v}=(\hbar/m)\partial_{\tilde{x}}
\mathrm{Arg}(\tilde{\Psi})$ is its local velocity.  Here and
throughout this paper, a tilde denotes a dimensional
quantity.

Non-dimensionalizing~\eqref{eq:unitsNLS} by scaling everything to
harmonic oscillator units
leads to the dimensionless or scaled NLSE,
\begin{multline}
  \label{eq:NLStime}
  \left[-\frac{1}{2}\frac{\partial^2}{\partial x^2} +
    g\vert\Psi(x,t)\vert^2 + V(x)\right]
  \Psi(x,t) \\=
  i\frac{\partial}{\partial t}\Psi(x,t),
\end{multline}
where
\begin{eqnarray}
  \label{eq:xnd}
  x &=& \frac{\tilde{x}}{\ell},\\
  \label{eq:tnd}
  t &=& \omega\tilde{t},\\
  \label{eq:gnd}
  g &=& \frac{\tilde{g}}{\hbar\omega\ell},\\
  \label{eq:Vnd}
  V &=& \frac{\tilde{V}}{\hbar\omega},\\
  \label{eq:psind}
  \Psi &=& \tilde{\Psi}\ell^{1/2},
\end{eqnarray}
and $\ell = \sqrt{\hbar/(m\omega)}$ is the harmonic oscillator length.

Equation~\eqref{eq:NLS} can
be solved~\cite{carr2000a} for stationary states of the form
\begin{equation}
  \label{eq:genansatz}
  \Psi(x,t)=\sqrt{\rho(x)}\,e^{i[\phi(x)-\mu t]},
\end{equation}
where $\mu$ is the eigenvalue of the stationary 1D NLSE.
Then the 1D NLSE becomes \begin{equation}
  \label{eq:NLS}
  \left[-\frac{1}{2}\frac{\partial^2}{\partial x^2} +
    g\vert\Psi(x,t)\vert^2 + V(x)\right]
  \Psi(x,t) =
  \mu \Psi(x,t).
\end{equation}
We work with the non-dimensionalized stationary 1D NLSE,
Eq.~\eqref{eq:NLS}, throughout the rest of our article.

\subsection{Exact Solution for a Constant Potential}
\label{ssec:exact}

When the potential $V(x)=V_0$ is a constant,
one can prove that the dimensionless density $\rho\equiv |\Psi|^2$
and the phase $\phi(x)$ have the form
\begin{eqnarray}
  \label{eq:gendens}
  \rho(x) &=& A\text{sn}^2\,(bx+\delta_0|m)+B,\\
  \label{eq:genphase}
  \frac{\partial\phi}{\partial x} &=& \frac{\alpha}{\rho},
\end{eqnarray}
where $A$ is the density scaling, $b$ is the translational
scaling, $m$ is the elliptic parameter, $\delta_0$ and $B$
are offsets, and $\alpha$ is an integration constant.  The
function $\text{sn}$ is one of twelve Jacobi elliptic
functions~\cite{abramowitz1964}, which can be interpreted
geometrically as the elliptical analog of the circular
trigonometric functions.  In this interpretation, the square
root of the elliptic parameter $m$ represents the
eccentricity of an ellipse.  In the limit that $m\to0$, the
Jacobi functions become the circular functions; in the limit
$m\to1$, the Jacobi functions become the hyperbolic
functions.

Relationships between parameters may be obtained by
substituting the solutions~\eqref{eq:gendens} and
\eqref{eq:genphase} into Eq.~\eqref{eq:NLS} and equating
coefficients of linearly independent powers of Jacobi
elliptic functions.  The relevant proofs of linear
independence are given in Appendix~\ref{app:proofs}.  We
obtain the relationships
\begin{eqnarray}
  m&=&\frac{A}{b^2}g,  \label{eq:meq}\\
  \mu &=& \frac{1}{2}[b^2+(A+3B)g]+V_0,  \label{eq:mueq}\\
  \alpha^2 &=& B(A+B)(b^2+Bg).
  \label{eq:alpha2eq}
\end{eqnarray}
Since $\alpha$ appears only in its square, and $\rho(x)$ does not
depend on $\alpha$, all solutions with $\alpha \neq 0$ are doubly
degenerate.  That is, $\pm\alpha$ result in solutions with the same
density $\rho(x)$ and eigenvalue $\mu$, but phases of opposite sign.

Mathematical and physical considerations lead us to conclude that not
all possible parameter values are relevant to this problem
\cite{RRMthesis}.  The relevant parameter space is summarized in Table
~\ref{tab:reparams}.

 \begin{table}[H]
   \centering
   \caption{Mathematical conditions on both NLSE parameters and solution parameters.}
    \begin{tabular}{l@{\hspace{0.7cm}}l}
      Parameter & Condition\\\hline
      $A$ & Real\\
      $B$ & Real\\
      $b$ & Either $\text{Re}(b) = 0$ or $\text{Im}(b) = 0$\\
      $\delta_0$ & No constraints\\
      $m$ & Real\\
      $g$ & Real\\
      $\mu$ & Real\\
      $\alpha$ & Real
    \end{tabular}
   \label{tab:reparams}
 \end{table}

\subsection{Calculation of Transmission}
\label{sec:Tcalc}

We consider a potential barrier of form
\begin{equation}
  \label{eq:potlbarr}
  V(x) =
  \begin{cases}
    0, \quad &x < x_1,\\
    V_0, \quad &x_1 < x < x_2,\\
    0, \quad &x_2 < x.
  \end{cases}
\end{equation}
We will take $V_0$ to be positive definite.  The barrier can
have any width and height, with the boundaries $x_1$ and
$x_2$ located at any positions along the $x$-axis.  We
define regions I and III to be the left and right sides of
the barrier, respectively, and region II as the region over
the barrier.  We apply our fully general solution from
Sec.~\ref{ssec:exact} to a numerical study of the
transmission of the BEC across the barrier.  We note that
the solution and our code can be generalized to arbitrary
piecewise-constant potentials with a finite number of jump
discontinuities.

Using the solution to the constant potential one can treat
any piecewise constant potential by the use of appropriate
boundary conditions, similar to the well-known method for
finding stationary solutions of the linear Schr\"odinger
equation with a rectangular well/barrier.  Our explicit
boundary conditions, expressed in terms of density and
phase, are
\begin{eqnarray}
\rho(x_i^-)&=&\rho(x_i^+),\\
\left.\frac{d\rho(x)}{dx}\right|_{x_i^-}&=&\left.\frac{d\rho(x)}{dx}\right|_{x_i^+},\\
\phi(x_i^-)&=&\phi(x_i^+)+2\pi n,\,n\in\mathbb{Z},\\
\alpha^-&=&\alpha^+\\
\mu^-&=&\mu^+
\label{eq:boundary}
\end{eqnarray}
where $x_i$ is the location of the $i$th boundary and $\pm$
indicates the right/left side of the
boundary~\cite{footnote2}.

The effective potential experienced by the BEC in the region
over the barrier is \begin{equation} \label{eq:muVeff}
V_{\rm eff} \equiv V_0 + g\rho(x).  \end{equation} In the
regime $\mu > \text{max}(V_{\rm eff})$, transmission over
the barrier is classically allowed.  For $\mu <
\text{max}(V_{\rm eff})$, transmission is classically
forbidden.  In both regimes, our scattering displays quantum
or wave-like effects, modified by the nonlinearity.  The
effective potential is why the linear and nonlinear problems
are so different, in addition to the lack of a superposition
principle.

We briefly remind the reader of the linear case.  For $g=0$
only in Eq.~\eqref{eq:NLS}, the wave function in each region
can be split into a sum of two terms representing left- and
right-traveling waves, using the principle of superposition.
We can take the left-hand side as the incident one.  On this
side, the solution contains both right- and left-traveling
waves, i.e., incident and reflected waves, with only
right-traveling waves on the transmission side of the
barrier.  In this case, we may define the transmission
coefficient as
\begin{equation}
  \label{eq:LinearTransmission}
  T =
  \frac{\langle\vert \Psi_{\text{trans}} \vert^2\rangle}{\langle\vert
    \Psi_{\text{inc}} \vert^2\rangle},
\end{equation}
where $\Psi_{\text{trans}}$ is the transmitted wave function and
$\Psi_{\text{inc}}$ is the incident wave function.  The angle brackets,
$\langle\hspace{1\jot}\cdot\hspace{1\jot}\rangle$, denote an average
value over one period of the function.  The definition given by
Eq.~\eqref{eq:LinearTransmission} is standard in linear quantum
mechanics.  In this interpretation, $T\leq1$ over the entire domain of
the system, and $T$ represents the probability that a given particle
will be transmitted across the barrier.

However, in the nonlinear case superposition does not apply.  We
cannot define separate left- and right-traveling waves in this case, and
thus the transmission coefficient is defined simply as
\begin{equation}
  \label{eq:NonlinearTransmission}
  T=\frac{\langle\vert\Psi_{III}\vert^2\rangle}{\langle\vert\Psi_{I}\vert^2\rangle} =\frac{\langle \rho_{III}\rangle}{\langle \rho_I\rangle},
\end{equation}
where $\Psi_{I}$ and $\Psi_{III}$ are the total wave
functions in regions I and III, with $\rho_{I}$ and
$\rho_{III}$ the corresponding number densities.  In
Eq.~\eqref{eq:LinearTransmission} this would be equivalent
to replacing the denominator with the sum of incident and
reflected amplitudes and then squaring to get the total
probability density to the left hand side of the barrier.
In the nonlinear case, since neither region I nor region III
can be said to be the incident side, we could just as easily
have replaced the definition in
Eq.~\eqref{eq:NonlinearTransmission} with its inverse.

Thus, in the nonlinear case, we find that $T$ may exceed
unity, according to Eq.~\eqref{eq:NonlinearTransmission}.
Physically, output cannot exceed input, and we conclude that
the ``transmission'' coefficient as we define it contains
information for atom lasers incident on either side of the
barrier.  Since we cannot choose only right- or
left-traveling waves in the solution, due to the
nonlinearity, we cannot define $T$ to restrict incidence to
only one side of the barrier, and must consider both
circumstances in the same solution set.

To calculate the average over the densities, we note that
the period of $\text{sn}^2\,(bx+\delta_0|m)$ is $2K(m)/b$,
where $K(m)$ is the complete elliptic integral of the first
kind~\cite{abramowitz1964}.  Thus we obtain the average
density by
\begin{equation}
  \label{eq:AvgDensIntegral}
  \langle \rho \rangle = \frac{b}{2K(m)} \int \!dx\,\rho(x),
\end{equation}
where the integral is taken over one period of $\rho(x)$.
Using the properties of Jacobi functions and elliptic integrals, one
can show that if $0\leq m \leq1$,
\begin{equation}
  \label{eq:AvgDensExplicit}
  \langle\rho\rangle = B + A\left[\frac{1}{m} - \frac{E(m)}{mK(m)}\right],
\end{equation}
where $E(m)$ is the complete elliptic integral of the second
kind.  If $m \notin [0,1]$, we must first apply a
transformation to write the density in terms of Jacobi
elliptic functions which depend on a parameter $m' \in
[0,1]$, then use the methods above to simplify the integral.
The relevant transformations are given in
Appendix~\ref{app:JEfns}.

\subsection{{Transmission Coefficients for Nonlinear
Scattering}}

As was mentioned previously, linear quantum mechanics
defines transmission as a ratio between the average of the
probability density function post-potential to that incident
on the potential. Moreover, this definition relies on the
linear decomposition of waves, which is not generally
available to nonlinear problems. Consequently, the
definition of transmission for nonlinear scattering problems
is not unique. In this section we briefly discuss our transmission
coefficient in comparison to another notable definition.

In~\cite{Paul2009} Paul \textit{et al.} consider  a mathematical
model, used to describe transport phenomenon of a BEC beam
that is
structurally similar to our Eq.~\eqref{eq:unitsNLS}. To
overcome the lack of superposition principles, the
authors adapt the methods of~\cite{Leboeuf2003,Paul2007} to
define a transmission coefficient valid in the limiting
regimes of
 \begin{enumerate}
  \item arbitrary nonlinearity and small transmission
	\item arbitrary transmission and small nonlinearity
 \end{enumerate}
where the decomposition of waves is approximately valid.
Experimentally, one considers a state where the
system is at rest with respect to the frame associated with
a travelling barrier. In this case, conservation of density
is recast into a first-order linear ordinary
differential equation (ODE) that when used with their
quasi-one-dimensional NLSE yields a nonlinear ODE on a
unitless amplitude, $A$, in the traveling variable,
$X=x-ct$, subject to an amplitude dependent potential,
$W(A)$. In upstream free-space, for when $X\to\infty$, the
first-integral of the previous ODE can be written as
 \begin{align}
   \frac{\hbar^{2}}{2m}\left(\frac{d\rho}{dX}\right)^{2} +
	 8F(\rho)
	 = 8\rho E^{+} \in \mathbb{R},
 \end{align}
which describes a fictitious classical particle with
``mass'' $\hbar^{2}/2m$, ``position'' $A=\sqrt{\rho}$ and ``time'' $X$,
evolving in a potential $W(A)=8F(\rho)$
\cite{Ivlev1984,Langer1967}.
Adapting the peturbative arguments of~\cite{Leboeuf2003}, the authors
power-expand $F$ about $\rho=1$ and find an ODE for the
perturbation $\delta \rho(X) = \rho(X)-1$ whose solution
gives $\rho(X) = 1 + 2\lambda + 2\sqrt{\lambda^{2}+\lambda}
\cos(2\kappa X + \theta)$, which describes upstream, $X\to
\infty$, density oscillations. Using the probability density,
$n$, defined by their NLSE and decomposing these oscillations into
incident and reflected waves gives,
\begin{align}
  \psi_{\text{inc}}(X)&=
	\sqrt{\frac{n(X)(1+\lambda)}{\rho(X)}}\mbox{exp}(-i\kappa
	X),\\
  \psi_{\text{ref}}(X)&=
	\sqrt{\frac{n(X)\lambda}{\rho(X)}}\mbox{exp}(i\kappa
	X+\theta),
 \end{align}
which, though only approximate solutions to the original
NLSE, gives an intuitive definition of transmission,
$T=1-|\psi_{\text{ref}}|^{2}/|\psi_{\text{inc}}|^{2}=(1+\lambda)^{-1}$.
It must be emphasized that, while intuitive, this definition is valid in
regime (1), $\lambda \ll 1$, and regime (2) when the
barrier speed is much larger than the BEC sound speed.
However, outside of these regimes where linear wave
decomposition cannot be applied, the dimensionless
parameter $\lambda$ finds continued use as the independent
variable of a Fokker-Plank equation whose solution defines
the probability distribution for transmission.

In conclusion, we have two definitions of nonlinear
transmission, both with valid
linear limits, see Appendix~\ref{app:convergence} Subsec.~\ref{subsec:linearlimits}. Regardless, either definition
corroborates the existence of grey soliton
trains and sinusoidal wake in experiments where the NLSE
with a
potential is the appropriate mathematical
model~\cite{Leboeuf2001}. It is interesting to note that the
nonlinearity presents itself as a balance between
calculation and intuition. That is, the work of Paul
\textit{et al.}
yields an intuitive transmission coefficient, which is
regime-limited by virtue of peturbative expansions.
On the other hand, our transmission coefficient is more general
but not as intuitive in the linear limit.

\subsection{Numerical Methods}

We use Mathematica to compute the transmission coefficient,
Eq.~\eqref{eq:NonlinearTransmission}.  We require an
internal precision of 100 digits; our main reason for using
Mathematica is its feature of arbitrary internal precision,
which is absolutely necessary for the highly singular
problem of nonlinear scattering.  Besides internal
precision, we use a uniform numerical tolerance of $10^{-5}$
for parameters.  If a quantity is smaller than this value,
it is taken to be zero; for example, we take $g < 10^{-5}$
to be zero.  We consider only repulsive interactions, i.e.,
$g \geq 0$, in our numerical analysis.  In addition, we
restrict analysis to positive barriers $V_0 > 0$, for
consistency with the idea of scattering by an atom laser.
However, as mentioned previously, our mathematics are
completely general, and our solution may be applied to
attractive interactions and potential wells.

To find a scattering solution, we take $A$, $B$, and
$\delta_0$ on the left side of the barrier, denoted by a
subscript $I$.  These parameters can be calculated uniquely
from the average number density, momentum, and energy for a
physical atom laser, as described in~\cite{carr2005a}.  We
consider the physical quantities $g$ and $\mu$, as well as
the barrier parameters, to be input parameters for an
experiment, and take them as known.  Other parameters on the
left side of the barrier may be obtained from
Eqs.~\eqref{eq:meq}--\eqref{eq:alpha2eq}.  We then use the
physical boundary conditions~\eqref{eq:boundary} to solve
for parameters in regions II and III in terms of the known
input parameters from region I.

Our code is completely symbolic, with the exception of one
numerical integration used in computing the transmission
coefficient.  In order to keep the code symbolic wherever
possible, we used Mathematica's pure function routines.
These are functions whose arguments are defined in terms of
their position rather than being given a specific variable
name.  This construction avoids the possibility of errors
arising from multiply-defined variables, while still
allowing us to maintain a consistent convention for
functional dependencies.

The analytical solution process using boundary conditions
yields twelve solutions for ($A$, $B$) in regions II and
III.  Six of these are extraneous solutions, obtained as a
result of squaring both sides of an equation.  Since they
are not true solutions, they do not satisfy the boundary
conditions, and this is used as a filter in the code to
discard these extraneous solutions.  The remaining six
solutions all correspond to the same functional form of the
density, and we need only consider one.  In our construction
of Eq.~\eqref{eq:gendens} as the polar form of a complex
function, we may take $\rho(x)$ to be real without loss of
generality.  Solutions with $\text{Im}(\rho)\neq0$ are
mathematically invalid, and these are encountered in the
code as a result of one or more assumptions breaking down.
Therefore, we take only real solutions for ($A$, $B$).

To obtain the average densities in
Eq.~\eqref{eq:NonlinearTransmission}, we numerically
integrate the densities using Simpson's
rule~\cite{press1993}.  The transmission coefficient is
computed as in \eqref{eq:NonlinearTransmission}, and is
treated as a resonance if it is within an interval of $\pm
10^{-4}$ around unity.

In the linear limit $g \to 0$, exact analytical expressions
may be obtained for the boundary conditions.  We use these
exact expressions, rather than Mathematica's Solve routine,
in the limit of small $g$.  By doing this, we avoid problems
with infinities due to internal Mathematica processes.

We also made use of several other methods to ensure the
correctness of the computations performed by Mathematica.
In our experience, Mathematica does not always handle
$\sqrt{-1}$ correctly.  Therefore, we kept $i$ as a symbol,
using replacement tables to handle powers of $i$, and
avoiding Mathematica's internal complex-number routines
wherever possible.  In addition, we included several
identities for Jacobi elliptic functions via replacement
tables, as Mathematica's default processes do not make use
of these identities for simplification.  Finally, rather
than relying on Mathematica to handle the special cases of
complex argument and $m>1$ internally, we used known
transformations to write equivalent expressions in terms of
Jacobi functions with real argument and parameter smaller
than unity.  These expressions were used in the code for
computations.

Convergence was verified using several methods.  These are detailed in
Appendix~\ref{app:convergence}.

\section{Soliton Localization}\label{sec:soliton}

 \begin{figure}[t]
   \begin{center}
   \subfloat[][]{\label{fig:densScatt}\includegraphics[width=0.4\textwidth]{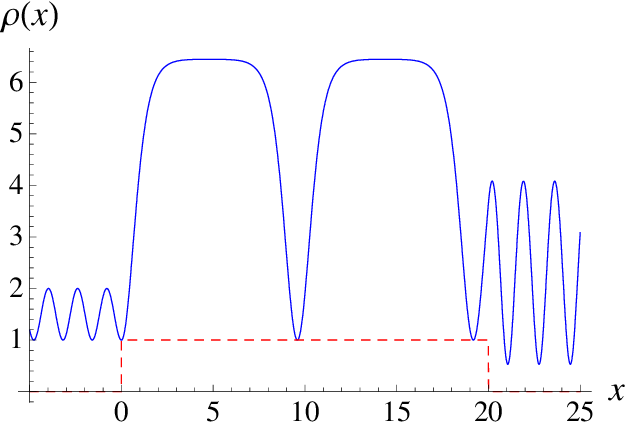}}\\
   \subfloat[][]{\label{fig:phaseScatt}\includegraphics[width=0.4\textwidth]{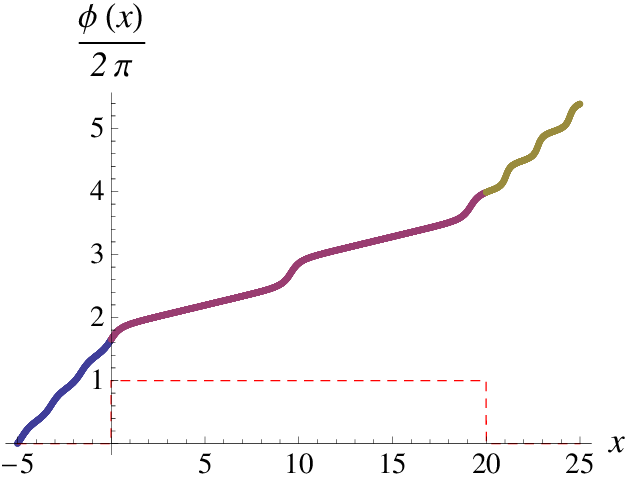}}
   \end{center}
\caption{(Color online) Density (upper panel) and phase
(lower panel) of a solution to the NLSE in which an incident
near-sinusoidal wave produces a single dark soliton
localized over the barrier. The density notch of a second
dark soliton is formed near the far right of the barrier,
around $x=19$, and a longer barrier could trap two or more
dark solitons in a dark soliton train.  The three colors in
the phase curve denote the three scattering regions.  The
red dashed curve shows the potential barrier for reference.}
\label{fig:densphaseScatt}
 \end{figure}

\begin{figure}[t]
   \centering
   \subfloat[][]{\label{fig:solitontraindens}\includegraphics[width=0.4\textwidth]{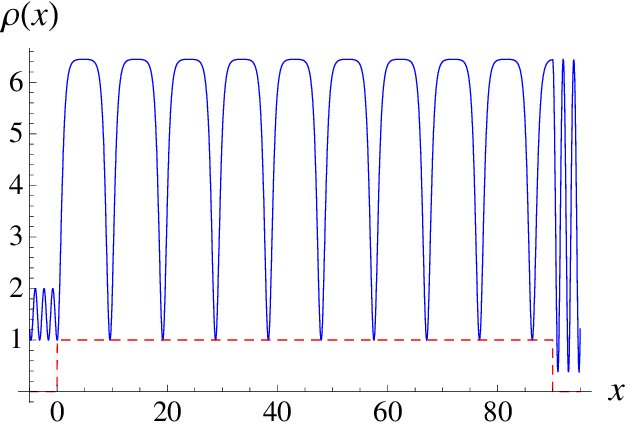}}\\
   \subfloat[][]{\label{fig:solitontrainphase}\includegraphics[width=0.4\textwidth]{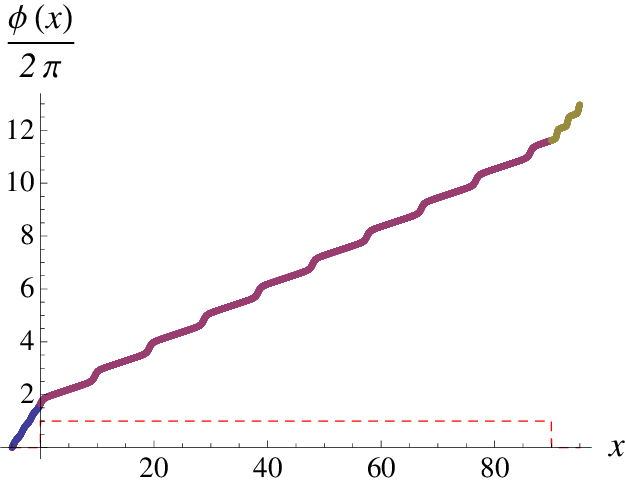}}
\caption{(Color online) Density (upper panel) and phase
(lower panel), similar to Fig.~\ref{fig:densphaseScatt}, but
with a soliton train localized on the barrier instead of a
single soliton.  The input parameters are identical to those
of Fig.~\ref{fig:densphaseScatt}; only the barrier width has
been changed, from 20 to 90.  We observe a train of 9 very
deep solitons localized on the barrier.}
\label{fig:solitontraindensphase} \end{figure}

A large variety of solution types appear after following the
methods of Sec.~\ref{sec:equation}.  We present here some
particularly interesting cases which are very far from the
kinds of solutions found for the linear Schr\"odinger
equation.  We recall that, in the linear Schr\"odinger
equation, the wavelength of the plane wave components across
regions I, II, and III is the same; it is only the spatial
offset in the arguments of the exponentials and the
amplitudes that can be different.  In the nonlinear case the
wavelength can be quite different in all three regions.  We
present two such cases, in which a phonon-like standing wave
is transformed into a strongly localized soliton or soliton
train over the barrier, and then returns to a phonon-like
profile.  We recall that phonons are a limit of Bogoliubov
quasi-particles for BECs, and appear as a small modulation
on top of a large offset, qualitatively speaking, as ripples
on a pond.  Since we work with the fully nonlinear problem
our phonons are not constrained to be small oscillations,
but continue to appear as ripples of amplitude $A$ and
translational offset $\delta_0$ on top of an offset of
magnitude $B$.

Figure~\ref{fig:densphaseScatt} shows the density and phase
for a localized soliton.  A well-localized solution can be
seen above the potential barrier.  The barrier, indicated
with a red dashed box in the figure, has height $V_0=1$ and
width 20.  The nonlinearity is $g=2.02$ and the chemical
potential $\mu=2.404$.  We choose $A_I=1$, $B_I=1$, and
$\delta_{0I}=0$.  In regions I and III, outside the barrier,
the elliptic parameter $m\ll 1$. In region II, over the
barrier, $m$ approaches unity and the peaks of the
$\text{sn}^2$ function broaden significantly.  The local
minimum of the density over the barrier is a dark soliton,
called ``grey'' because it does not have a node.  Such a
soliton is always moving.  Examining the phase profile, we
observe two areas in region II: away from the dark soliton
is a background superflow with a shallower slope, while over
the soliton there is a characteristic phase jump displaying
a sharp increase in slope.  In this solution the superflow
flows to the right with velocity $v=\partial_x \phi(x)$,
while the dark soliton moves to the left with an equal and
opposite velocity; the two velocities cancel each other,
leading to a stationary state.

The phase has an overall linear envelope on either side of
the barrier, with clearly visible oscillations on this
background.  The background slopes differ slightly between
regions I and III, and the slope over the barrier in region
II is smaller by a factor of 3.  This shows that the
velocity profile of the BEC need not be the same on either
side of the barrier.  Although we fixed parameters on the
left-hand side in order to find this solution, as described
in Sec.~\ref{sec:equation}, from the physical perspective of
an atom laser the BEC is incident on the right-hand side.
Thus, as happens approximately half the time in our solution
method, we have in fact fixed the output rather than the
input parameters.  The overall transmission right to left is
about 50\%, and $T>1$ according to our definition from
Eq.~\eqref{eq:NonlinearTransmission}.

With a longer barrier of width $x_2-x_1=90$ and the same nonlinearity $g=2.02$, chemical potential $\mu=2.404$, and input parameters $A_I=1$, $B_I=1$, and $\delta_{0I}=0$ fixed on the left hand side, we obtain a soliton train localized on the barrier, as shown in Fig.~\ref{fig:solitontraindensphase}.  In Fig.~\ref{fig:solitontraindensphase} we again display the potential with a dashed red line for reference.  The location of the dark solitons can be observed in both the density and the phase, similar to Fig.~\ref{fig:densphaseScatt}.  Multiple solitons have been observed in a variety of BEC experiments~\cite{strecker2002,engelsP2007,weller2008}.  However, such solitons have so far not been phase-locked into a train in BECs.  Our barrier method constitutes a totally new way of producing multiple solitons locked into a train.  The length of the barrier can be used to control the number of solitons in the train.  We note that dark soliton trains have been produced in nonlinear spin waves in thin magnetic films~\cite{kalinikos2000,wuMingzhong2004}, but not via our scattering technique; also spin waves are in fact a damped driven system which are best modeled by an open-system version of the NLSE quite different from Eq.~\eqref{eq:NLS}~\cite{carr2010g}.


\section{Atom Laser Transmission}\label{sec:transmission}

We turn now to a more systematic exploration of the solution
space.  The nonlinear Schr\"odinger equation has a
substantially larger parameter space to explore than the
linear case.  We divide this exploration into wide and
narrow barriers.  The barrier size $x_2-x_1$ can be compared
to the BEC healing length, $\xi \equiv
1/\sqrt{8\pi\langle\rho\rangle a_s}$, with
$\langle\rho\rangle$ the average linear number density.
However, the healing length only provides a useful
comparison for low-energy excitations, as it describes how a
uniform ground-state BEC is perturbed by a localized
potential, e.g. a hard wall.  This corresponds to
well-separated dark solitons, as in region II in
Figs.~\ref{fig:densphaseScatt}
and~\ref{fig:solitontraindensphase}.  As we treat a wide
variety of excitations, many of which take standing-wave
form which is very far from the ground state, e.g. regions I
and III in Figs.~\ref{fig:densphaseScatt}
and~\ref{fig:solitontraindensphase}, the correct comparison
to identify wide and narrow barriers is in fact the
wavelength of the excitations, \begin{equation}
\lambda\equiv \frac{2K(m)}{b} =
\frac{2K\left(\frac{Ag}{2(\mu-V_0)-(A+3B)g}\right)}{\sqrt{2(\mu-V_0)-(A+3B)g}},
\label{eq:wavelength} \end{equation} as can be calculated
from Eqs.~\eqref{eq:meq} and~\eqref{eq:gendens}.  Thus
$x_2-x_1 \ll \lambda_j$ is the narrow barrier case and
$x_2-x_1 \gg \lambda_j$ is the wide barrier case, where the
subscript $j\in\{I,II,III\}$ refers to the region.  For
sufficiently small $\lambda$ all barriers are effectively
wide.  Holding other parameters fixed, including the
nonlinearity, the wide barrier limit occurs for large
chemical potential, since the wavelength is a decreasing
function of $\mu$, as can be verified by consideration of
Eq.~\eqref{eq:wavelength}.

We found that the barrier height $V_0$ is less important, a
statement which we will support further in
Sec.~\ref{sec:resonances}.  This dependence on width more
than height is another sense in which the nonlinear case is
very different from the linear case; in the latter it is
only the effective area of the barrier,
$2M\tilde{V}_0(\tilde{x}_2-\tilde{x}_1)^2/\hbar^2$, that is
important.  The dependence on width arises mainly from the
fact that the density in the nonlinear case can have
different wavelengths in different regions for the same
solution, in contrast to the linear case, as previously
mentioned.

For simplicity we will keep the input parameters $A_I$,
$B_I$, and $\delta_{0I}$ fixed to the same values as those
used in
Figs.~\ref{fig:densphaseScatt}-\ref{fig:solitontraindensphase}
and Sec.~\ref{sec:soliton}; after a massive exploration of
the parameter space, we found all results to be
qualitatively similar to those described below.

\subsection{Wide Barrier Case}\label{sec:wideV}

 \begin{figure}[t]
   \centering
  \includegraphics[width=0.4\textwidth]{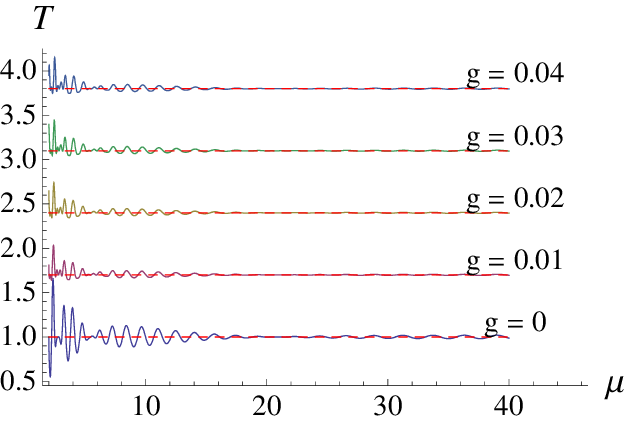}
\caption{(Color online) \emph{Wide barrier, small
nonlinearity}.  Transmission for wide barrier with
increasing nonlinearity from bottom to top, starting with
the linear case, $g=0$, in steps of $\Delta g = 0.01$. Even
a very small nonlinearity changes the problem because the
left-right symmetry of the effective potential is broken.}
\label{fig:wideBarrierSmallNonlinearity} \end{figure}

 \begin{figure}[t]
   \centering
   \includegraphics[width=0.4\textwidth]{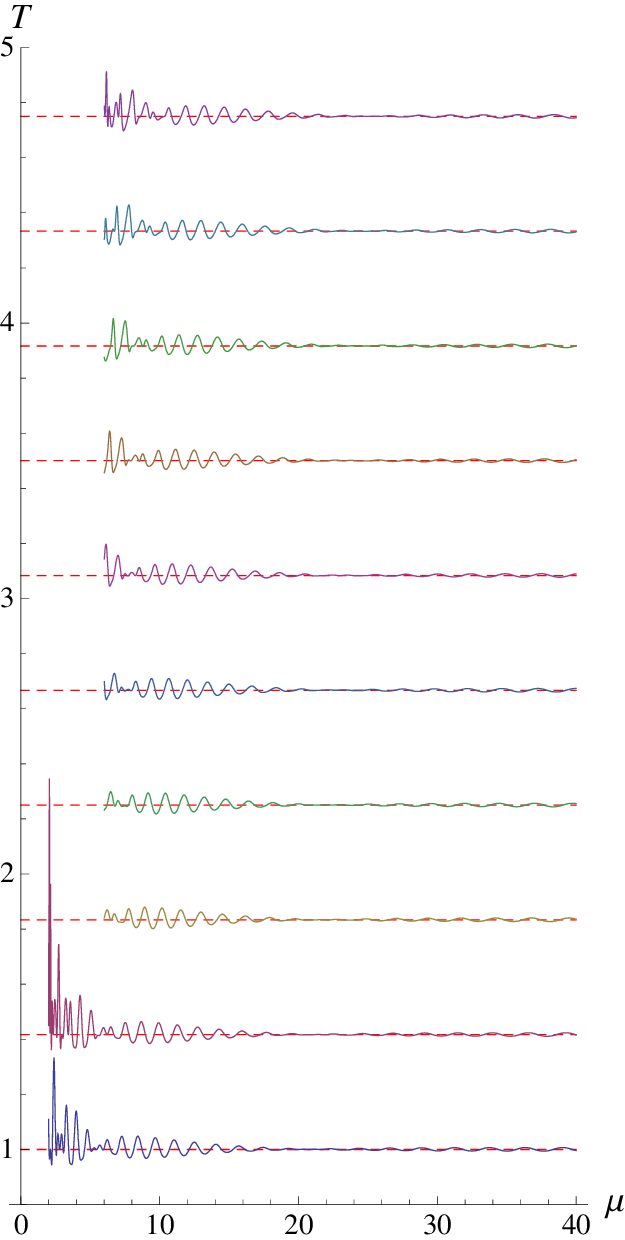}
\caption{(Color online) \emph{Wide barrier, medium
nonlinearity}.  Same as
Fig.~\ref{fig:wideBarrierSmallNonlinearity}, but with steps
of $\Delta g =0.1$, from $g=0.1$ to $g=1.0$  An extended
region of near-perfect invisibility of the barrier in the
$\mu\simeq 20$ to $\mu\simeq 30$ range translates to the
right as the nonlinearity increases.}
\label{fig:wideBarrierMediumNonlinearity} \end{figure}

We consider a barrier whose width is much larger than its
height.  We take a barrier of width $x_2-x_1=20$ and height
$V_0=1$ for the purposes of illustration.  In
Figs.~\ref{fig:wideBarrierSmallNonlinearity}-\ref{fig:wideBarrierMediumNonlinearity}
transmission plots are laid out on the same set of
horizontal axes.  The dashed red lines denote $T=1$, and
solid curves are the transmission coefficient as defined in
Eq.~\eqref{eq:NonlinearTransmission}.  The nonlinearity $g$
increases in steps of 0.01 in
Fig.~\ref{fig:wideBarrierMediumNonlinearity} as we move
upward on the plot, and in steps of 0.1 in
\ref{fig:wideBarrierSmallNonlinearity}.  Each transmission
plot is at a convenient vertical offset for illustration,
but the plots are not otherwise scaled or shifted.  The
nonlinearity is shown next to each curve for reference.  As
discussed in Section~\ref{sec:Tcalc}, the transmission
coefficient may be greater than 1.  This is not a novel
physical feature; it arises due to the redefinition of
transmission as in Eq.~\eqref{eq:NonlinearTransmission} and
the invalidity of superposition in this problem.

Comparing the linear, i.e $g=0$, transmission curve in Fig.
\ref{fig:wideBarrierSmallNonlinearity} with the other
transmission plots in the same figure, we observe that there
is a significant change in transmission behavior when we
enter the nonlinear regime, even for very small
nonlinearity, $g\sim \mathcal{O}(10^{-2})$.  The most
significant change occurs when $\mu$ is small, meaning that
the potential barrier has a greater overall effect on the
condensate.  In the linear case, $T$ oscillates about evenly
on either side of unity for small $\mu$, and the amplitude
of oscillations is similar in either direction.  In the
nonlinear case, we see that while transmission can still be
less than unity, it does not drop as far below unity as in
the linear case.  We understand this difference to be due to
the effective potential: for $g=0$ the operator in the NLSE
is not biased, whereas for non-zero $g$ it is.  Thus, by
choosing parameters on the left hand side, we fix the
effective potential and bias the system towards particular
parameter sets.

As $g$ increases, new peaks appear in the small-$\mu$
regime, and the behavior of $T$ changes significantly for
smaller $\mu$.  This regime does not appear for $g \gtrsim
0.2$.  The reason that the transmission curves depend
strongly on $g$ in the small-$\mu$ regime is that $g$ and
$\mu$ are not completely independent.  For a given chemical
potential $\mu$, when the nonlinearity $g$ becomes larger
than a certain cut-off value, solutions are generated that
have $\rho(x) \in \mathbb{C}$.  This is invalid because,
when we solved the NLSE, we assumed $\rho(x) \in
\mathbb{R}$.  We may take $\rho(x) \in \mathbb{R}$ without
loss of generality, since Eq.~\eqref{eq:genansatz} is a
general polar representation of a complex number.  For large
$\mu$ the cut-off in $g$ is pushed to a region off the top
of our plot; see figures in Sec.~\ref{sec:resonances} for
more details.  The appearance of complex $\rho(x)$ means
that one or more of our assumptions are breaking down in
this regime.  A more detailed analysis of this situation is
considered in~\cite{RRMthesis}.  One such break-down is due
to the appearance of quasi-bound states with complex
$\mu$~\cite{carr2004b,carr2005c,dekel2009}.

For larger values of $\mu$, the overall behavior of the
transmission does not change significantly between each
plot.  However, we do see a shift in the transmission curve.
For higher $\mu$, the transmission plot retains its shape
and shifts to the right as $g$ increases.  This feature is
especially apparent when a computer is used to animate the
transmission plots for increasing $g$; in
Fig.~\ref{fig:wideBarrierMediumNonlinearity} we have
endeavored to lay such an animation out on the page.  In
Fig.~\ref{fig:wideBarrierMediumNonlinearity}, the regime of
the same curve-shape translating gradually to the right with
increasing $g$ occurs when $\mu \gtrsim 6$.

Another feature of note is that with the definition
\eqref{eq:NonlinearTransmission} of transmission, the
amplitude of oscillations in $T$ does not decrease
monotonically as in the usual linear interpretation.  In all
of the transmission plots of
Fig.~\ref{fig:wideBarrierMediumNonlinearity}, we see
significant oscillations on either side of a transition
region between $\mu\simeq 20$ and $\mu \simeq 30$.  In this
region, we have almost perfect resonance, i.e., $T$ is very
close to unity.  Further analysis of transmission resonances
will be presented in Sec.~\ref{sec:resonances}.

\subsection{Narrow barrier}\label{sec:narrowV}

 \begin{figure}[!htb]
   \centering
   \includegraphics[width=0.4\textwidth]{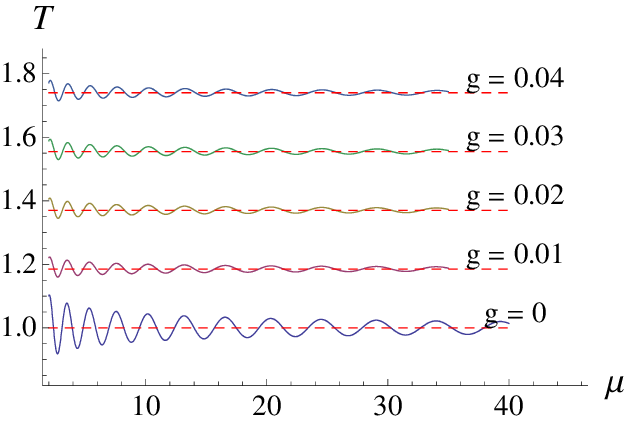}
   \caption{Transmission for narrow barrier with small nonlinearity.
     Note the difference in behavior between the
     linear and nonlinear regimes.}
   \label{fig:narrowBarrierSmallNonlinearity}
 \end{figure}

 \begin{figure}[!htb]
   \centering
\includegraphics[width=0.4\textwidth]{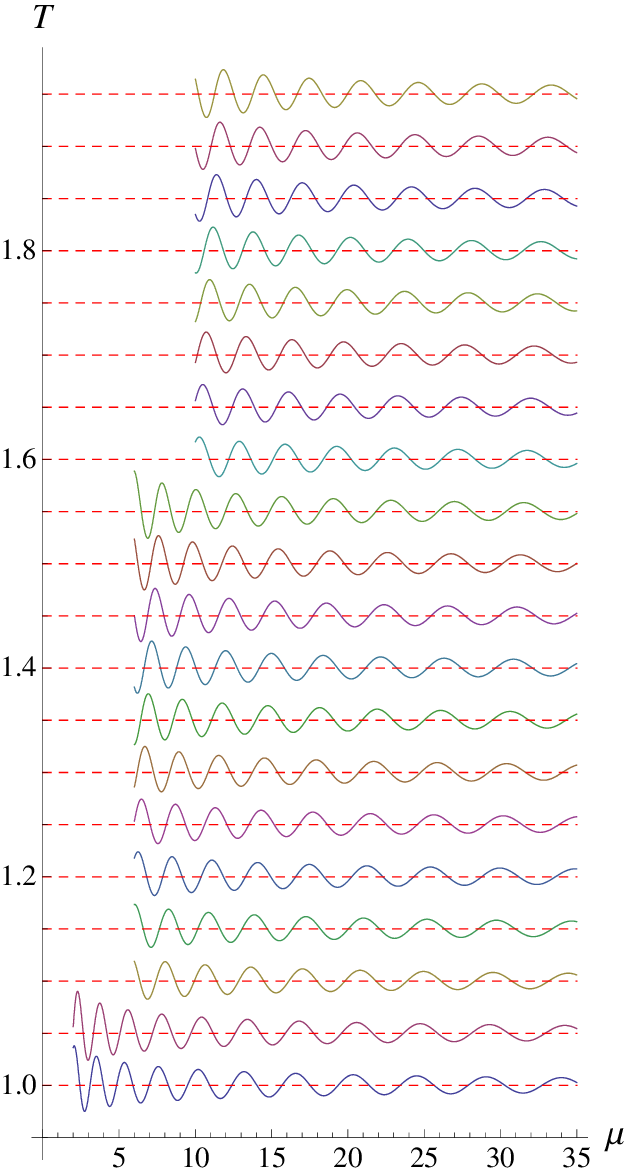}
   \caption{Transmission for narrow barrier with increasing
     nonlinearity from bottom to top, from $g=0.1$ to $g=2.0$ in steps of $\Delta g = 0.1$.}
   \label{fig:narrowBarrierMediumNonlinearity}
 \end{figure}

To explore the narrow-barrier regime, we choose
$x_2-x_1=0.1$ and $V_0=10$, so that the barrier is a factor
of ten narrower than it is high; note that the height is the
same as the wide-barrier case explored in
Sec.~\ref{sec:wideV}.  Over the full range of values of
$\mu$ and $g$ that we consider, the narrow-barrier
requirement, $\lambda \gg x_x-x_1$, is satisfied, as can be
verified from Eq.~\eqref{eq:mueq}; in the range we consider,
$\lambda \geq 0.35$.  We first focus on transmission plots
for small nonlinearity, in
Fig.~\ref{fig:narrowBarrierSmallNonlinearity}.  The layout
of the figure is the same as in Sec.~\ref{sec:wideV}: five
transmission plots are shown on the same set of horizontal
axes, with vertical offsets in $T$ for illustration, the
plots are not otherwise shifted or scaled, and $g$ is
increased from zero in steps of $\Delta g = 0.01$.
Comparing the linear $g=0$ transmission curve with those for
$g>0$, we again see a significant drop in the amplitude of
oscillations as we go from the linear to the nonlinear
regime.  Again, the nonlinear transmission curves tend to
stay further above unity than below unity, although the
difference is less pronounced than in the wide-barrier case.
As a function of $\mu$ the transmission curve is much
smoother as compared to
Fig.~\ref{fig:wideBarrierSmallNonlinearity}.  This is
because there is always less than one wavelength fitting
into the barrier, and thus a small change in $\mu$ cannot
suddenly cause an integer number of wavelengths to match the
barrier width.

In Fig.~\ref{fig:narrowBarrierMediumNonlinearity}, we turn
to the regime of medium nonlinearity, increasing $g$ from
0.1 in steps of $\Delta g = 0.1$ up to $g=2.0$.  The
amplitude of oscillations decreases smoothly for increasing
$\mu$, independent of $g$.  There is no region of
near-perfect resonance as in the wide-barrier case: $T$
oscillates quasi-periodically about $T=1$.  In fact, the
entire transmission curve simply translates smoothly to the
right as $g$ increases, with no real abrupt behavior even
for the smallest values of $\mu$ allowed for each curve.
Thus we can surmise that the abrupt behavior in
Sec.~\ref{sec:wideV} is due to the barrier width, and it is
the width, not the height, that mainly controls the
transmission curves.  In the following section we will
adduce further evidence to support this point.

\section{Transmission Resonances and Bifurcations}\label{sec:resonances}

\begin{figure}[t]
  \centering
  \includegraphics[width=0.4\textwidth]{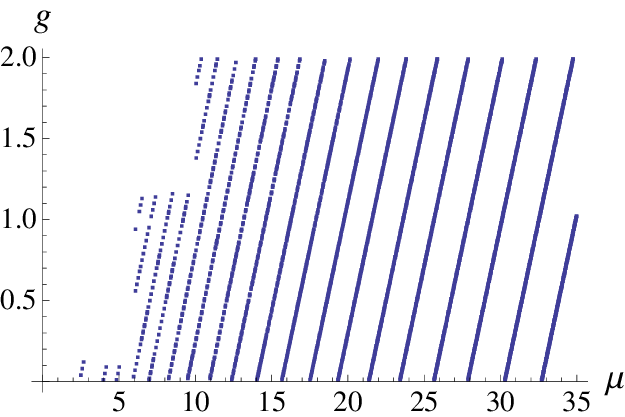}
	\caption{Transmission resonances for narrow barrier.  All
	behavior is described by regularly spaced straight lines
	except for very small $\mu$, where some combinations of
	$g$ and $\mu$ do not produce stationary solutions.}
	\label{fig:res1}
\end{figure}

To better understand the results of
Sec.~\ref{sec:transmission}, in particular the extended
region of near-perfect invisibility of the barrier, we focus
on the transmission resonances only, i.e., the points in the
$g$--$\mu$ plane for which $T=1$ to within a tolerance of
$10^{-5}$.

We first treat the simpler case of the narrow barrier.  The
transmission resonances from
Fig.~\ref{fig:narrowBarrierMediumNonlinearity} are shown in
Fig.~\ref{fig:res1}.  This clearly shows that all
transmission resonances translate smoothly to the right for
increasing $g$, with a constant slope.  This figure allows
us to predict the location of all transmission resonances
except for small $\mu$, where constraints in the allowed
simultaneous values of $g$ and $\mu$ cut off certain
regions, as discussed in Sec.~\ref{sec:transmission}.  We
also observe in Fig.~\ref{fig:res1} that the spacing between
curves increases with increasing $\mu$. A functional fit to
this spacing can allow us to predict transmission resonances
in the entire $g$--$\mu$ plane.  Since the spacing is the
same starting from $g=0$, it is straightforward to find such
a function for any values of $A$, $B$, and $\delta_0$.

\begin{figure}[t]
  \centering
  \includegraphics[width=0.4\textwidth]{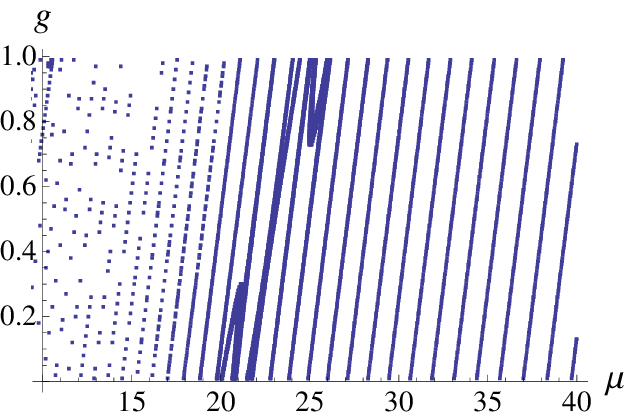}\\
  \caption{\label{fig:res2}Transmission resonances for wide barrier. Bifurcations occur in an extended region of near-perfect invisibility of the barrier.}
\end{figure}

With the narrow-barrier case as a reference, we move on to
the more intricate wide-barrier case.  Figure~\ref{fig:res2}
shows the transmission resonances from
Fig.~\ref{fig:wideBarrierMediumNonlinearity}.  We observe
that the resonances are sparse for low values of $\mu$, with
much more sporadic and abrupt behavior than the narrow
barrier case.  This is because an integer number of
wavelengths can fit into the barrier width, an effect which
is more pronounced for small $\mu$; for instance, for
$g=0.02$ and $\mu=2$, $\lambda=2.27$, while for $\mu=40$,
$\lambda=0.355$.  In the regions where lines of resonance
appear, the spacing between these lines is smaller than in
Fig.~\ref{fig:res1}.  The spacing increases as $\mu$
increases, though not as fast as in the narrow-barrier case.

The resonances are sparse for lower values of $\mu$ and more
uniform for higher values of $\mu$.  For mid-range $\mu$, we
see very different behavior.  There are three regions
between $\mu=20$ and $\mu=26$ where the resonances are
extremely dense.  These correspond to the region of
near-constant resonance seen in the transmission plots.
This region shifts to the right as $g$ increases, as
observed in the transmission plots.  Three bifurcations are
present in this region.  Bifurcations are typical of
nonlinear systems; the particular example we show here for
this parameter set is expected to be a generic feature for
all parameter sets.  Further numerical exploration found
these three bifurcations continued in an apparently infinite
sequence sloping away upwards to the right.  We found that a
linear function was not sufficient to fit such bifurcations,
and they occurred near but not precisely at a value of
$\lambda = 1/2$.  We did not find bifurcations at other
rational values of $\lambda$, so this may be a coincidence;
the barrier length in this parameter set is $x_2-x_1 = 10$.
We explored up to $\mu=70$ and $g=10$.  Although our
computationally intensive study was only performed
thoroughly for this particular choice of parameters, we did
spot checks through many regions of parameter space and
observed similar extended regions of near-perfect
invisibility of the barrier.

\begin{figure}[H]
  \centering
  \includegraphics[width=0.4\textwidth]{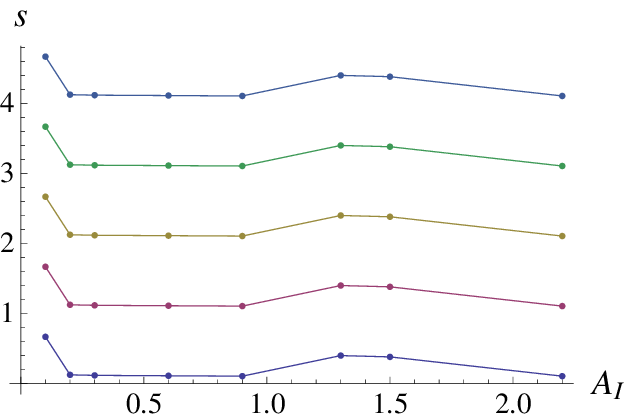}
	\caption{Slopes of resonance transmission lines, as a
	function of density offset, coarse sampling.  The curves
	show $B=0.2$, 0.5, 1.0, 1.5, 2.0 from bottom to top; a
	vertical offset is given to each curve for visualization,
	since they in fact all lie on top of each other.  The
	points represent actual data, while the curves are a guide
	to the eye.} \label{fig:coarseslopeplot}
\end{figure}

\begin{figure}[H]
  \centering
  \includegraphics[width=0.4\textwidth]{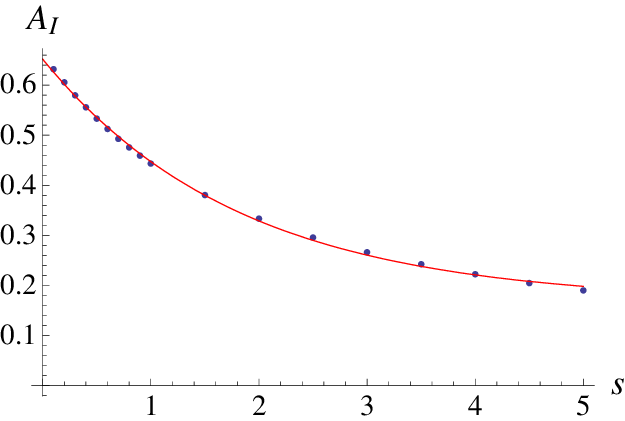}
	\caption{Slope of resonance curves as a function of input
	amplitude.  Although the transmission resonances curves do
	not depend on $B$ and $\delta_0$, they do depend strongly
	on $A$.  Points are numerical calculations, while the
	solid curve is a least squares fit.}
	\label{fig:slopeplot2}
\end{figure}

We return to the narrow barrier case for further analysis.
We consider the slope $s$ of these parallel lines of
resonance as a function of other physical parameters.  For
the parameter regimes considered in this study, the slope of
the resonance lines does not depend on the value of the
density offset, $B$, as seen in
Fig.~\ref{fig:coarseslopeplot}.  In this plot, the slope is
shown for several values of $B$.  Each curve has been
vertically shifted by a convenient offset for illustration;
all curves in fact lie on top of each other.  We note that
the elliptic parameter $m$, which is strongly governed by
the nonlinearity of the system, depends on both $A$ and $b$,
but does not depend on $B$, as described in
Sec.~\ref{ssec:exact}.  Similarly, we found that the slopes
$s$ do not depend on the spatial translational offset
$\delta_0$.  However, the slopes do depend strong on the
amplitude $A$.  The slope decreases as the amplitude of the
input density increases. We find an exponential
least-squares fit as shown in Fig.~\ref{fig:slopeplot2}:
\begin{equation} \label{eq:fiteqn} s(A_I) = 0.167 + 0.485
e^{-0.549 A_I}, \end{equation} where we fixed $B_I=1.0$.

\section{Conclusions}\label{sec:conclusions}

We developed a general method for obtaining stationary
states of the nonlinear Schr\"odinger equation for any
piecewise constant potential.  We applied this method to
nonlinear scattering on a rectangular potential barrier,
focusing on cnoidal waves, or dark soliton trains.  This
problem differs greatly from the textbook linear quantum
mechanics problem of scattering on a rectangular barrier.
Among novel nonlinear features are the following.  First,
the wavelength in the three regions (left of the barrier, on
the barrier, and right of the barrier) need not be the same.
As a consequence, such a barrier can be used to create one
or more sharply localized dark solitons from a broad
phonon-like input. Second, it is mainly the barrier width,
not its area, that controls the kind of stationary states
observed. For wide barriers in which the barrier width is
larger than the wavelength extended regions of near-perfect
transmission occur; the barrier is invisible over a range of
interaction strength $g$ and chemical potential $\mu$.
Third, in wide barriers an apparently infinite sequence of
bifurcations appears in this invisibility region.  Despite
the parameter space for solutions being large, we showed
that it is just the amplitude that determines slopes of
transmission resonance lines, greatly simplifying this
complex problem.

To relate our predictions to alternate physical input of
average density, momentum, and energy of an atom laser
formed from a Bose-Einstein condensate, the analysis laid
out in Ref.~\cite{carr2005a} may be used directly; thus we
do not repeat it here.  Future possibilities for this work
include applying our method to various piecewise constant
potentials of interest in applications, and a more detailed
mathematical study of the bifurcations we found in
transmission resonances.

We acknowledge helpful discussions with Paul A. Martin and
Matthew Heller.  {This material is based in part upon
work supported by the National Science Foundation under
Grant Numbers PHY-0547845 and PHY-1067973.  L.D.C. thanks
the Alexander von Humboldt foundation for additional
support, and the Institute for Advanced Study at Tsinghua
University for hosting him during completion of a
significant portion of this work.}

\appendix

\section{Jacobi Elliptic Functions}\label{app:JEfns}
There are twelve Jacobi elliptic functions in all: sn, cn, dn, sc, nc,
dc, cs, ds, ns~\cite{abramowitz1964}.  Not all of these are independent.  The Jacobi functions are doubly-periodic
in the complex plane and they depend on two parameters: an independent
variable $u$ and the \emph{elliptic parameter} $m$.  For real parameter $m$,
we may assume $0 \leq m \leq 1$.  If $m$ is outside of this range,
transformations can be made to write the Jacobi function in terms of
functions whose parameter is between 0 and 1.  For
$\text{sn}\,(u\vert m)$,
which appears in the density $\rho(x)$ of the BEC, these
transformations are, for $m<0$,
\begin{eqnarray}
  \label{eq:negmsntrans}
  \text{sn}\,(u\vert m) &=& \left(
\frac{1}{1-m}\right)^{1/2}\nonumber\\
&&\times\,\text{sd}\left[(1-m)^{1/2}u\bigg\vert\left(\frac{-m}{1-m}\right)\right],
\end{eqnarray}
and for $m>1$,
\begin{equation}
  \label{eq:bigmtrans}
  \text{sn}\,(u\vert m) = m^{-1/2}\text{sn}\,(u m^{1/2}|m^{-1}).
\end{equation}

When $m$ lies between zero and unity, the Jacobi functions
can be interpreted geometrically as the analog of the
hyperbolic and trigonometric functions.  In this
interpretation, the parameter $m$ corresponds to the
eccentricity of the ellipse.  For $m=0$, the Jacobi
functions reduce to the trigonometric functions; for $m=1$,
to the hyperbolic functions.

The Jacobi elliptic functions are defined as inverse
integrals~\cite{bowman1}, and by the locations of zeros and
poles in the complex plane \cite{abramowitz1964}.  They may
be related to one another by various
identities~\cite{fettis1972,khare2002,dayton2002}.

\section{Proofs}\label{app:proofs}

\subsection{Linear Independence of Powers of $\bm{\text{sn}\,(u|m)}$}\label{sec:linind}

In the following we state a particularly vital proof which
we have not found elsewhere in the literature, and is
required to establish our exact solutions.  Other proofs can
found in Ref.~\cite{RRMthesis}.  \theorem
\label{thm:linindsn} The functions $\text{sn}^p(u|m)$ and
$\text{sn}^q(u|m),~p,q \in \mathbb{Z}$, are linearly
independent for $p \neq q$.

\proof
Compute the Wronskian of the two functions:
\begin{align}
  &W[\text{sn}^p(u|m),\text{sn}^q(u|m)] =\nonumber\\
  &\qquad\qquad\begin{vmatrix}
    \text{sn}^p(u|m) & \text{sn}^q(u|m)\\
    \frac{\partial}{\partial u}[\text{sn}^p(u|m)] &
    \frac{\partial}{\partial u}[\text{sn}^q(u|m)]
  \end{vmatrix}\\
  \nonumber
  &= q\text{sn}^p(u|m)\text{sn}^{q-1}(u|m)\text{cn}\,(u|m)\text{dn}\,(u|m)\\
  & \qquad - p \text{sn}^{p-1}(u|m)\text{sn}^q(u|m)\text{cn}\,(u|m)\text{dn}\,(u|m)\\
  &=(q-p)\text{cn}\,(u|m)\text{dn}\,(u|m)\text{sn}^{p+q-1}(u|m).
\end{align}
The product of Jacobi elliptic functions is nonvanishing except on a
set of measure zero; therefore, for $p \neq q$ the functions are
linearly independent.  QED.

\subsection{Linear Independence of Products of Powers of $\bm{\text{sn}\,(u|m)}$}\label{sec:linindprod}
Let
\begin{align}
  &f_1(x) = \text{sn}^p(a|m)\text{sn}^q(u|m),\\
  &f_2(x) = \text{sn}^r(a|m)\text{sn}^s(u|m),
\end{align}
where $p$, $q$, $r$, $s$ are integers, and $a$ and $u$ are linear
functions of $x$.  We may assume without loss of
generality that $a\neq u$.  The case $a=u$ is analyzed in
Section~\ref{sec:linind}.  Computing the Wronskian of $f_1, f_2$, we find
\begin{multline}
  \label{eq:prodwronsk}
  W(f_1,f_2) =\\
  (r-p)\text{cn}\,(a|m)\text{dn}\,(a|m)\text{sn}^{r+p-1}(a|m)\text{sn}^{s+q}(u|m) +\\
  (q-s)\text{cn}\,(u|m)\text{dn}\,(u|m)\text{sn}^{s+q-1}(u|m)\text{sn}^{r+p}(a|m).
\end{multline}
The Wronskian vanishes $\forall a, u$ only when $r=p$ and $q=s$; that
is, when $f_1$ and $f_2$ are not distinct functions.  In all other
cases, $f_1$ and $f_2$ are linearly independent.

\section{Numerical Convergence}\label{app:convergence}

We consider several limits of the NLSE,
Eq.~\eqref{eq:NLS}.  These are used to verify consistency of our code with known results.

\subsection{Linear Limit}\label{subsec:linearlimits}

\begin{figure}[H]
  \centering
  \subfloat[][]{\label{fig:NonLinTLinLimit}\includegraphics[width=0.4\textwidth]{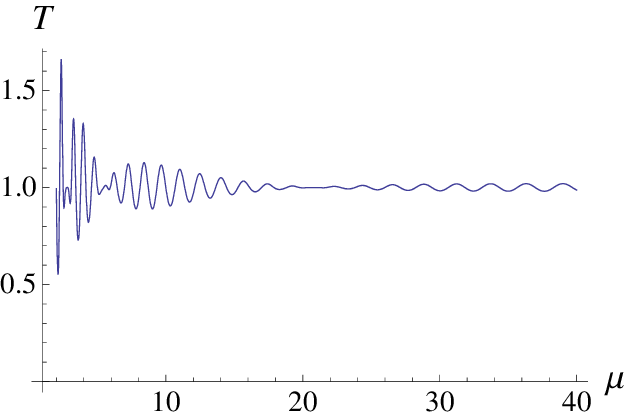}}\\
  \subfloat[][]{\label{fig:LinTLinLimit}\includegraphics[width=0.4\textwidth]{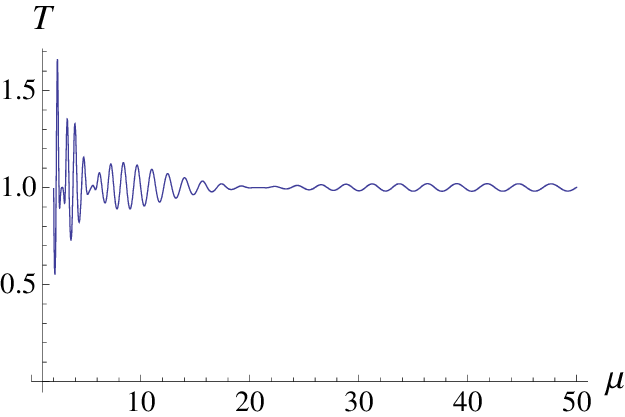}}\\
  \subfloat[][]{\label{fig:LinNonLinErr}\includegraphics[width=0.4\textwidth]{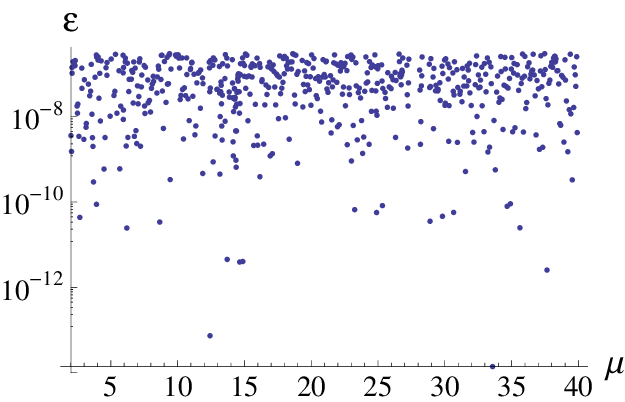}}
  \caption{(a) Numerically computed transmission for linear limit.  (b) Exact linear value of transmission.  (c) Error in transmission for linear limit.}
  \label{fig:LinNonLinComp}
\end{figure}

We consider the NLSE, Eq.~\eqref{eq:NLS}, in the limit that $g
\to 0$.  In this limit, the solution should reduce to the
well-known linear scattering solution, which is analyzed in
many elementary quantum mechanics texts.  In the linear
limit, we find that $m \to 0$, so that the linear density is
\begin{equation}
  \label{eq:lindens}
  \rho(x) = A\sin^2(bx + \delta_0),
\end{equation}
since $B=0$ in the linear case~\cite{RRMthesis}.  For
comparison with the nonlinear case, we define transmission
as $\langle \rho_{III} \rangle/ \langle \rho_I \rangle$.
The integral for $\langle \rho \rangle$ can be evaluated
exactly in this case:
\begin{align}
  \label{eq:rhointlinear}
  \langle \rho \rangle &= \int_0^{2\pi/b} \!dx\, A\sin^2(bx + \delta_0)\,,
  \\
  &= \tfrac{1}{2} A.
\end{align}
Therefore, the transmission is
\begin{equation}
  \label{eq:Tlinlimit}
  T_{\ell} = \frac{A_{III}}{A_I}.
\end{equation}
We can compare the value given by Eq.~\eqref{eq:Tlinlimit} to the
value $T$ obtained numerically by the code.  Transmission plots are shown
in Figs.~\ref{fig:NonLinTLinLimit} and~\ref{fig:LinTLinLimit}.   A
log plot of the error,
\begin{equation}
  \label{eq:linearTerr}
  \varepsilon = \frac{\vert T-T_{\ell}\vert}{T_{\text{avg}}},
\end{equation}
is given in Fig.
\ref{fig:LinNonLinErr}.  The maximum value of the error is
$\mathcal{O}(10^{-7})$, which is within the numerical tolerance of the
code.

\subsection{Constant Potential Limit}\label{subsec:zeroV}
\begin{figure}[!htb]
  \centering
  \includegraphics[width=0.4\textwidth]{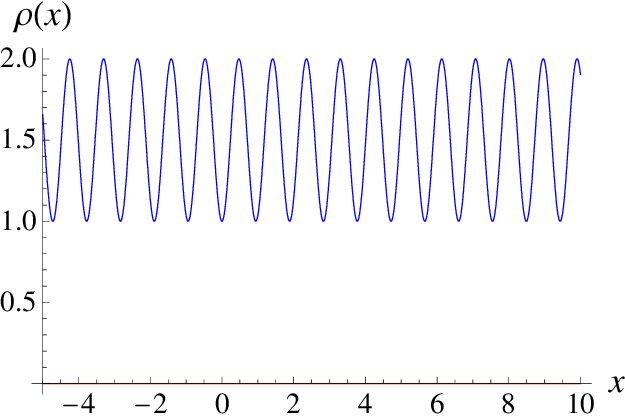}
  \caption{Density plot for zero barrier.}
  \label{fig:rhoPlotZeroV1}
\end{figure}

Consider the NLSE, Eq.~\eqref{eq:NLS}, with potential barrier
\eqref{eq:potlbarr}, in the limit that $V_0 \to 0$.  In this case,
boundary conditions are redundant and we expect all parameters to be
constant $\forall x$.  By setting a ``barrier'' of $V_0 = 0$ in the
code, we can verify that the code gives the correct solution; namely,
the amplitude, period, and shifts in the density should not change at
the ``boundary'' locations.  Indeed this is the case, providing an
additional verification of correctness for the code.  A density plot
for this case, with a ``barrier'' of width 5 and height 1, is shown in
Fig.~\ref{fig:rhoPlotZeroV1}.

\begin{figure}[!htb]
  \centering
  \includegraphics[width=0.4\textwidth]{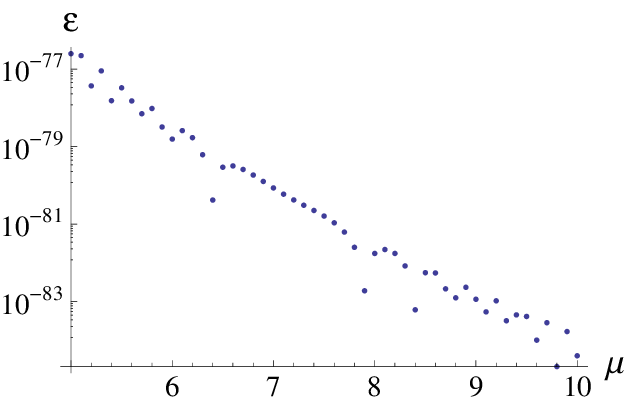}
  \caption{Error in transmission for zero barrier.}
  \label{fig:errorPlotZeroV}
\end{figure}

Since the density parameters do not change over space, we expect to
find a transmission coefficient of 1.  We compute and plot the error
\begin{equation}
  \label{eq:zeroVerr}
  \varepsilon = \ln\left(\frac{\vert T-1\vert}{T_{\text{avg}}}\right),
\end{equation}
where $T_{\text{avg}}$ denotes the average value of
transmission over the plot interval.  A log plot of the error is given
in Fig.~\ref{fig:errorPlotZeroV}.

\begin{figure}[!htb]
  \centering
  \includegraphics[width=0.4\textwidth]{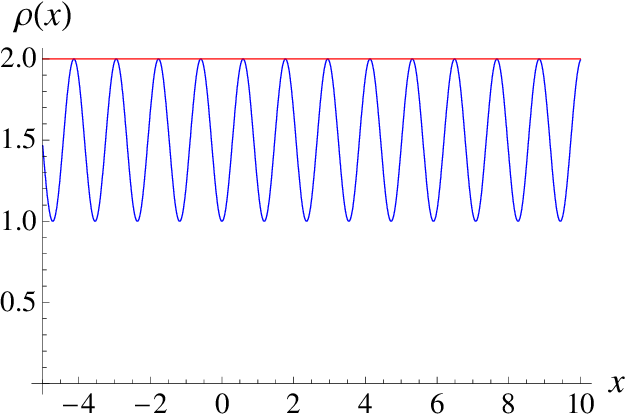}
  \caption{Density plot for nonzero flat barrier.}
  \label{fig:rhoPlotConstV1}
\end{figure}

Alternatively, we can consider a constant nonzero potential:
$V(x) = V_c,~\forall x$ in Eq.~\eqref{eq:NLS}.  Again we
expect all parameters to be constant $\forall x$.  We set a
``barrier'' of $V_I = V_{II} = V_{III} = 2$ and width 5 in
the code and plot the density.  The plot is shown in Fig.
\ref{fig:rhoPlotConstV1}.

\begin{figure}[!htb]
  \centering
  \includegraphics[width=0.4\textwidth]{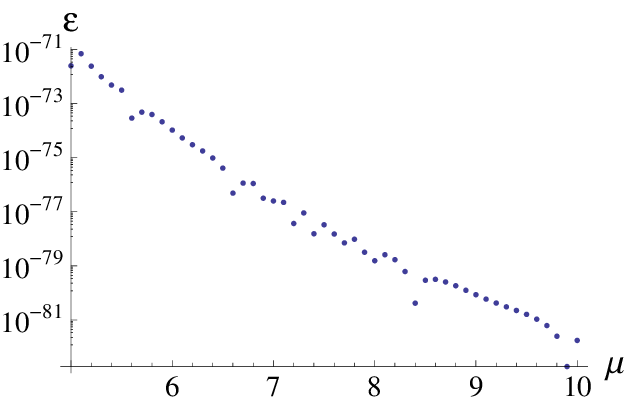}
  \caption{Error in transmission for nonzero flat barrier.}
  \label{fig:errorPlotConstV}
\end{figure}

Again, since the density parameters do not change over
space, we expect to find a transmission coefficient of 1.
We compute and plot the error
\begin{equation}
  \label{eq:constVerr}
  \varepsilon = \ln\left(\frac{\vert T-1\vert}{T_{\text{avg}}}\right),
\end{equation}
where $T_{\text{avg}}$ denotes the average value of
transmission over the plot interval.  A log plot of the error is given
in Fig.~\ref{fig:errorPlotConstV}.

Therefore the code gives the expected results for density
and transmission in the constant-potential limits.

\subsection{Thomas-Fermi Limit}
We consider the limit that $\hbar \partial^2\Psi/\partial x^2\to0$ in the unscaled NLSE,
Eq.~\eqref{eq:unitsNLS}.  In this limit, the unscaled NLSE becomes
\begin{equation}
  \label{eq:NLSclassical}
  \left[ g\vert\Psi(x,t)\vert^2 + V(x)\right] \Psi(x,t) = i\hbar
  \frac{\partial}{\partial t}\Psi(x,t).
\end{equation}
Substituting Eq.~\eqref{eq:genansatz} for the wave function
$\Psi$ in~\eqref{eq:NLSclassical} and rearranging, we obtain
\begin{equation}
  \label{eq:NLSclassSub}
  g\rho^{3/2} + [V(x)-\hbar\mu]\rho^{1/2} = 0,
\end{equation}
so that either $\rho(x) \equiv 0$, the trivial case, or else
\begin{equation}
  \label{eq:rhoclassical}
  \rho(x) = [\hbar\mu - V(x)]/g,
\end{equation}
when $g\neq0$.  Note that Eq.~\eqref{eq:rhoclassical} works for any
spatially-dependent potential $V(x)$.

Equation~\eqref{eq:rhoclassical} is the well-known
Thomas-Fermi limit~\cite{dalfovo1999}.  It is relevant when
the curvature of $\Psi$ is nearly zero.  This can be
accomplished either by taking the limit $A/B\to0$, so that
the amplitude of oscillations is small, or by taking the
limit $1/b \to \infty$, so that the wavelength of
oscillations is large.


\begin{thebibliography}{10}
\bibitem{footnote1}
The results of our study can be qualitatively applied to a barrier of smooth
  shape with the same area as the rectangular barrier.
\bibitem{footnote2}
The final condition, $\mu^-=\mu^+$, is not strictly required when the density
  vanishes on the boundary; however, in the latter case, one can argue this
  condition holds on physical grounds~\cite{RRMthesis}.
\end{thebibliography}

\end{document}